\documentstyle[12pt]{article}
\begin{document}
\topmargin 0pt
\oddsidemargin -3.5mm
\headheight 0pt
\topskip 0mm
\addtolength{\baselineskip}{0.20\baselineskip}
\begin{flushright}
SOGANG-HEP $190/94$
\end{flushright}
\vspace{1.5cm}
\begin{center}
{\Large \bf  Operator Ordering Problem of the Nonrelativistic Chern-Simons
Theory }
\end{center}
\vspace{0.5cm}
\begin{center}
{Mu-In Park$^{a}$ and Young-Jai Park$^{b}$}\\
\vspace{0.5cm}
{Department of Physics and Basic Science Research Institute, }\\
{Sogang University, C.P.O.Box 1142, Seoul 100-611, Korea}\\
\vspace{1.5cm}
{\large \bf ABSTRACT}
\end{center}

The operator ordering problem due to the quantization or regularization
ambiguity in the Chern-Simons theory exists. However, we show that this can
be avoided
if we require Galilei covariance of the nonrelativistic Abelian
Chern-Simons theory even at the quantum level for the extended sources.
The covariance can be recovered only by choosing some particular operator
orderings for the generators of the Galilei
group depending on the quantization
ambiguities of the $gauge-matter$ commutation relation. We show that the
desired ordering for the unusual prescription is not the same as the
well-known normal ordering but still satisfies all the necessary conditions.
Furthermore, we show that the equations of motion can be
expressed in a similar form regardless of the regularization ambiguity. This
suggests that the different regularization prescriptions do not change the
physics. On the other hand, for the case of point sources the
regularization prescription is uniquely determined, and only the orderings,
which are equivalent to the usual one, are allowed.
\vspace{2.5cm}
\begin{flushleft}
\end{flushleft}
\newpage
\begin{center}
{\large \bf I. INTRODUCTION } \\
\end{center}

Recently, the Chern-Simons gauge theory has captured considerable
interest due to the fact that it is still the only known example of
Galilei-covariant gauge theory [1], or it can realize Wilczek's charge-flux
composite model of the anyon [2,3]. Furthermore, it has been proposed
as a toy
model
of quark confinement in $D=2+1$ [4]. It has been also extensively studied
as the
topologically massive gauge theory with the Maxwell term [5]. But quantal
analysis
of the Chern-Simons theory [1,3,6,7] shows that there is the operator
ordering problem due
to the quantization or regularization ambiguity in defining the quantization
rule, i.e., commutation relation of gauge-matter. Although they
chose a particular regularization prescription such that there is no operator
ordering problem, it is still unclear whether one can avoid the
ordering problem
even for more general regularization prescriptions, or determine the correct
prescriptions from some first principles [3].

In this paper we show that the ordering problem can be avoided, and the key
resides in an unusual property of the nonrelativistic Chern-Simons gauge
theory. In contrast with our common expectation, the Galilei covariance of the
nonrelativistic Chern-Simons gauge theory in the Galilei-covariant gauges is
not a trivial matter. The covariance can be only recovered by choosing some
particular operator orderings for the generators of the Galilei group. The
extended and point source systems are separately analyzed. Since the point
source systems with more general regularization prescriptions need much more
care than the extended ones, we will first treat the extended case. In Sec. II
the Galilei covariance of the nonrelativistic Abelian Chern-Simons
gauge theory
in the Coulomb gauge is examined for the extended sources. We explicitly show
that the covariance can be recovered only by choosing some particular operator
orderings depending on the quantization ambiguities of the gauge-matter
commutation relations. Moreover, we point out that the desired ordering
for the
unusual prescription is not the same as the well-known normal ordering but
still satisfies all the necessary conditions, while the desired ordering
for the
usual prescription is the same as the usual one. In Sec. III the operator
equations of motion for the properly ordered generators for the extended
sources
are examined. As a result, all the equations of motion corresponding
to different
orderings can be expressed in a similar form. In Sec. IV the ordering problem
for the point source system is examined, and we show that the usual
prescription
is the only possible one. Hence the orderings, which are equivalent
to the usual
one, are only allowed. Section V is devoted to summary and remarks
about several
generalizations of our analysis.
\\

\begin{center}
{\large \bf II. GALILEI COVARIANCE IN COULOMB GAUGE} \\
\end{center}

The nonrelativistic Abelian Chern-Simons gauge theory on the plane is described
by the Lagrangian
\begin{eqnarray}
{\cal L}=\frac{\kappa}{2} \epsilon^{\mu \nu \rho}A_{\mu}\partial_{\nu}A_{\rho}
+i\phi^{*}D_{t}\phi-\frac{1}{2m}(D_{k}\phi)^{*}(D_{k}\phi),
\end{eqnarray}
where $\epsilon^{012}=1$, $g_{\mu \nu}$=diag(1,-1,-1), and $D_{\mu}=
\partial_{\mu}+iA_{\mu}$. It is invariant up to the total divergence under the
gauge transformations
\begin{eqnarray}
\phi \rightarrow exp[-i\Lambda] \phi,~~A_{\mu} \rightarrow A_{\mu} +
\partial_{\mu}\Lambda,
\end{eqnarray}
where $\Lambda$ is a well-behaved function such that
$\epsilon^{\mu \nu \lambda}
\partial_{\mu} \partial_{\nu} \Lambda=0$. Then the classical equations
of motion
are
\begin{eqnarray}
B &\equiv& \epsilon^{ij} \partial_{i}A^{j}
             = -\frac{1}{\kappa} J_{0},   \\
E^{i} &\equiv&  -\partial_{i} A^{0}-\partial_{t}A^{i} =
             \frac{1}{\kappa} \epsilon^{ij}J^{j},
\end{eqnarray}
where $J_{0}=\phi^{*}\phi,~ J^{i}=(1/2mi) \{ \phi^{*}D_{i}\phi
 - (D_{i}\phi^{*})\phi \}$, and
 $\epsilon^{0ij} \equiv \epsilon^{ij}$.
We will quantize this model by defining usual equal-time
commutation relations for the matter field  as
\begin{eqnarray}
&&[ \phi({\bf x}), \phi^{\dagger}({\bf x'}) ]
                = \delta^{2}({\bf x-x'}), \nonumber \\
&&[ \phi({\bf x}), \phi({\bf x'})]  = 0 ,~
{}~[ \phi^{\dagger}({\bf x}), \phi^{\dagger}({\bf x'}) ] = 0
\end{eqnarray}
by considering $\phi$ to be a quantum field and $\phi^{*}$ its dagger
$\phi^{\dagger}$. Note that although several authors have considered the usual
Dirac
procedure [8] or symplectic quantization method [9], we consider here
the safest
approach in order not to miss the ordering problem following Refs.
[1], [3], [4], and
[7]. In this approach one adopts quantum rules such as (5),
which do not possess
the ordering ambiguities as the fundamental ones.

Now let us consider Eqs. (3) and(4) as the operator equations. If we choose the
Coulomb gauge, the solutions for ${\bf A}$ and $A^{0}$ are given by [1,7]
\begin{eqnarray}
{\bf A}({\bf x}, t)=\overline{ {\bf \nabla}} \frac{1}{\kappa}
  \int d^{2}x' {\cal{D}}({\bf x-x'}) J_{0}({\bf x'},t), \\
A^{0}({\bf x},t)=-\frac{1}{\kappa} \int d^{2} x'{\bf J}({\bf x'},t)
  \cdot \overline{{\nabla}}' \cal{D}({\bf x-x'}),
\end{eqnarray}
where $\overline{\nabla}^{i}=\epsilon^{ij} \partial_{j}$, having the property
$\overline{\nabla} ^{2}=\nabla ^{2}$, and $\cal{D}(\bf{x})$ is defined by
\begin{eqnarray}
{\nabla}^{2} {\cal D}({\bf x}) = {\delta}^{2} ({\bf x})
\end{eqnarray}
and has the well-known solution
\begin{eqnarray}
{\cal D}({\bf x})=\frac{1}{4 \pi} ln{\bf x}^{2},
\end{eqnarray}
by neglecting the trivial constant term. Then, the gauge potentials
$\bf{A}$ and
$A^{0}$ become the quantum operators due to the operator nature of the
densities
$J_{0}, ~\bf{J}$. Although solutions (6) and (7) are generally valid for both
point and extended sources, we first consider the extended one in this section.
Moreover, since all the gauge fields $\bf{ A}$ and $A^{0}$ are completely
expressed by the matter fields, we may expect that the commutation relations of
the gauge-gauge and gauge-matter should be completely determined by the
matter-matter commutation relation (5). But as was pointed out in Refs. [6] and
[7], there is quantization ambiguity in the gauge-matter commutation
relation at
the same point
\begin{eqnarray}
[{\bf A}({\bf x}), \phi({\bf x})] =
\left[-\frac{1}{\kappa}\overline {\bf \nabla} {\cal D}({\bf x-x'})
\mid_{\bf x=x'}\right]\phi({\bf x}),
\end{eqnarray}
since $ \overline {\nabla}^{i} {\cal D}({\bf x}) = ( 1/2\pi) \epsilon^{ij}
x^{j} / {\bf x}^{2} $ is ill defined at the origin.
Although several authors [1,6,7], by choosing a particular regularization
preserving the antisymmetry of $\overline{\bf{\nabla}} \cal{D}$ under space
reflection, have assumed
$\overline {\bf{\nabla}} \cal{D}({\bf{x-x'}}) {\mid}_{\bf{ x=x'}}$ = 0
such that
there is no operator ordering problem, it is still unclear whether the
ordering problem can be also avoided for even more general regularization
prescriptions, or whether the correct prescriptions can be determined
from some first
principles [3].

In the following, we present a novel feature of the nonrelativistic Abelian
Chern-Simons theory, which is that the Galilei covariance in the
Galilei-covariant
gauges, like the Coulomb gauge, is spoiled for incorrect orderings depending
on the regularization prescriptions. To this end, we first write Eq. (10) as
\begin{eqnarray}
[{\bf A}({\bf x}), \phi({\bf x})] ={\bf F}\phi({\bf x}),
\end{eqnarray}
where $\bf{F}$ denotes
$ -(1/\kappa)\overline {\bf{\nabla}}\cal{D}(\bf{x-x'})\mid_{\bf{x=x'}}$, which
is real and should be constant in both space and time such that the
quantization rule (11) is covariant under space and time translation [10].
Then,
there exists the nontrivial ordering ambiguity in defining the covariant
derivatives $\bf{D \phi}$ and $\bf{D \phi^{\dagger}}$; i.e., we can define
covariant derivatives as
\begin{eqnarray}
&&{\bf D}^{(1)} \phi({\bf x}) \equiv  {\bf \nabla}\phi({\bf x})
    -i {\bf A}({\bf x}) \phi({\bf x}), \nonumber \\
&&{\bf D}^{(1)} \phi^{\dagger}({\bf x}) \equiv {\bf \nabla}\phi^{\dagger}
  ({\bf x})-i {\bf A}({\bf x}) \phi^{\dagger}({\bf x}), \nonumber  \\
&&{\bf D}^{(2)} \phi({\bf x}) \equiv {\bf \nabla}\phi({\bf x})
           -i \phi({\bf x}){\bf A}({\bf x}), \nonumber \\
&&{\bf D}^{(2)} \phi^{\dagger}({\bf x}) \equiv  {\bf \nabla}\phi^{\dagger}
  ({\bf x})- i \phi^{\dagger}({\bf x}){\bf A}({\bf x}).
\end{eqnarray}
Then, the differences between these two different definitions
${\bf D}^{(1)}\phi$(or $\phi^{\dagger}$) and ${\bf D}^{(2)}\phi$(
or $\phi^{\dagger}$) are given by the field-dependent terms as
\begin{eqnarray}
&&[{\bf D}^{(1)} -{\bf D}^{(2)}] \phi({\bf x})
                =-i{\bf F} \phi({\bf x}), \nonumber \\
&&[{\bf D}^{(1)} -{\bf D}^{(2)}] \phi^{\dagger}({\bf x})
                =-i{\bf F} \phi^{\dagger}({\bf x}).
\end{eqnarray}
As a result, one can easily expect that these produce the nontrivial operator
ordering problem to the generators of the current operator $\bf{J}$ and Galilei
group $H$ (time translation), $\bf{P}$ (space translation),
$\bf{G}$ (Galilei boost), and $J$ (rotation).

\begin{center}
\bf{A. Hamiltonian operator}
\end{center}

The classical Hamiltonian in our analysis is given by
\begin{eqnarray}
H &=&\int d^{2}x T^{00} \nonumber \\
  &=&\int d^{2}x \frac{1}{2m}[{\bf D} \phi({\bf x})]^{*}\cdot
      [{\bf D} \phi({\bf x})] \nonumber \\
  &=&\int d^{2}x \frac{1}{2m}[{\bf \nabla}\phi({\bf x})-
     i {\bf A}({\bf x}) \phi({\bf x})]^{*}\cdot
     [{\bf \nabla}\phi({\bf x})-i {\bf A}({\bf x}) \phi({\bf x})],
\end{eqnarray}
using the nonrelativistic energy momentum tensor $T^{\mu \nu}$ [7]
\begin{eqnarray*}
&&T^{00}=\frac{1}{2m}( D_{i} \phi)^{*}  ( D_{i} \phi),  \\
&&T^{0i}=\frac{1}{2i}[ \phi^{*} D_{i} \phi-(D_{i} \phi )^{*} \phi ],  \\
&&T^{i0}=-\frac{1}{2m}[(D_{t} \phi)^{*}(D_{i} \phi)
     + (D_{t} \phi)(D_{i} \phi)^{*}], \\
&&T^{ij}=\frac{1}{2m}[(D_{i} \phi)^{*}(D_{j} \phi)+ (D_{i} \phi)(D_{j}
\phi)^{*}
  -\delta_{ij} (D_{k} \phi)^{*}(D_{k} \phi)]  \\
&&~~~~~~~~  +\frac{1}{4m}(\delta_{ij} \nabla^{2}
           -2 \partial_{i} \partial_{j} )J_{0}+\delta_{ij} T^{00}.
\end{eqnarray*}

Then, four different Hermitian operator forms of the Hamiltonian are possible:
\begin{eqnarray}
H_{a} &\equiv& \int d^{2}x \frac{1}{2m}[{\bf D}^{(1)} \phi({\bf x})]^{\dagger}
  \cdot [{\bf D}^{(1)} \phi({\bf x})] \nonumber \\
  &=&\int d^{2}x \frac{1}{2m}[{\bf \nabla}\phi({\bf x})-
     i {\bf A}({\bf x}) \phi({\bf x})]^{\dagger}\cdot
     [{\bf \nabla}\phi({\bf x})-i {\bf A}({\bf x}) \phi({\bf x})],\nonumber \\
H_{b} &\equiv& \int d^{2}x \frac{1}{2m}[{\bf D}^{(2)} \phi({\bf x})]^{\dagger}
  \cdot [{\bf D}^{(2)} \phi({\bf x})] \nonumber \\
      &=& H_{a}+{\bf F} \cdot \int d^{2} x {\bf J}^{(1)}-
      \frac{1}{2m} {\bf F}^{2} Q, \nonumber \\
H_{c} &\equiv& \int d^{2}x \frac{1}{2m}[{\bf D}^{(1)} \phi({\bf x})]
  \cdot [{\bf D}^{(1)} \phi({\bf x})]^{\dagger} \nonumber \\
   &=& H_{b} + \frac{\delta ^{2} (0)}{2m} \int d^{2} x {\bf A}^{2}({\bf x}),
   \nonumber \\
H_{d} &\equiv& \int d^{2}x \frac{1}{2m}[{\bf D}^{(2)} \phi({\bf x})]
  \cdot [{\bf D}^{(2)} \phi({\bf x})]^{\dagger} \nonumber \\
   &=& H_{c} + {\bf F} \cdot \int d^{2} x {\bf J}^{(1)}
   -\frac{1}{2m} {\bf F}^{2} Q
   -\frac{\delta ^{2} (0)}{m} {\bf F} \cdot \int d^{2} x {\bf A}
   + \frac{{\bf F}^{2}}{2m} V,
\end{eqnarray}
where ${\bf J } ^{(1)} = (1/2mi) [ \phi^{\dagger}({\bf D}^{(1)} \phi)
       - ({\bf D}^{(1)}\phi)^{\dagger} \phi ]$,
$\delta^{2} (0)=\delta^{2}(\bf{x-x})$, $Q=\int d^{2} x J^{0}$, and
$ V=\int d^{2} x $. The corresponding quantum field equations of motion for the
matter field are given by
\begin{eqnarray}
&&(i \partial_{t} \phi({\bf x}) )_{a}  \equiv  [\phi({\bf x}), H_{a}]
     \nonumber \\
  &&~~~ =- \frac{1}{2m} { {\bf D}^{(1)} }^{2} \phi({\bf x})
    +{A^{0}}^{(1)}({\bf x}) \phi({\bf x})
   +  \frac{1}{2m {\kappa}^{2} }  \int d^{2} x'
    [\overline {\bf \nabla}' {\cal D}({\bf x'-x})]^{2} J_{0}({\bf x'})
     \phi({\bf x}),\nonumber \\
&&(i \partial_{t} \phi({\bf x}) )_{b} \equiv  [\phi({\bf x}), H_{b}]
     \nonumber \\
   &&~~~= [ \phi({\bf x}), H_{a} ] - \frac{i}{m} {\bf F} \cdot {\bf D}^{(1)}
   \phi({\bf x})
    + \frac{1}{2m} {\bf F}^{2} \phi({\bf x})
    - \frac{1}{m \kappa } {\bf F} \cdot \int d^{2} x'
    [\overline {\bf \nabla}' {\cal D}({\bf x'-x})] \nonumber \\
   &&~~~ \times J_{0}({\bf x'}) \phi({\bf x}), \nonumber \\
&&(i \partial_{t} \phi({\bf x}) )_{c} \equiv [\phi({\bf x}), H_{c}] \nonumber\\
   &&~~~ = [\phi({\bf x}), H_{b}]
    + \frac{\delta^{2} (0)}{m \kappa } \int d^{2} x'
    [\overline {\bf \nabla}' {\cal D}({\bf x'-x})] \cdot {\bf A}({\bf x'})
     \phi({\bf x})
    + \frac{\delta^{2} (0)}{2m {\kappa}^{2} } \int d^{2} x'
    [\overline {\bf \nabla}' {\cal D}({\bf x'-x})]^{2} \nonumber \\
    &&~~~ \times \phi({\bf x}), \nonumber \\
&&(i \partial_{t} \phi({\bf x}) )_{d} \equiv  [\phi({\bf x}), H_{d}]
    \nonumber \\
   &&~~~ = [\phi({\bf x}), H_{c}] -\frac{i}{m} {\bf F} \cdot {\bf D}^{(1)}
     \phi({\bf x})
    + \frac{1}{2m} {\bf F}^{2} \phi({\bf x})
    - \frac{1}{m \kappa } {\bf F} \cdot \int d^{2} x'
    [\overline {\bf \nabla}' {\cal D}({\bf x'-x})] J_{0}({\bf x'})
     \phi({\bf x})  \nonumber \\
    &&~~~- \frac{\delta^{2} (0)}{m \kappa } {\bf F} \cdot \int d^{2} x'
    [\overline {\bf \nabla}' {\cal D}({\bf x'-x})] \phi({\bf x}) ,
\end{eqnarray}
where the scalar potential ${A^{0}}^{(1)}$ is
\begin{eqnarray}
{A^{0}}^{(1)}({\bf x},t)=-\frac{1}{\kappa} \int d^{2} x'{\bf J}^{(1)}
      ({\bf x'},t)
  \cdot \overline{{\nabla}}' \cal{D}({\bf x-x'}).
\end{eqnarray}
Note that it is clear from the expression of ${\bf  J}^{(1)}$ that there is
also
quantization ambiguity in an $A^{0}$-matter commutation relation. This problem
is related to the nonuniqueness of the current operator and will be treated
later. At
present, the ${\bf{A}}$-matter commutation relation is only needed in
calculating the generators of the Galilei group. Moreover, we note that the
last
term in the first equation of (16) is the usual Jackiw-Pi quantum correction
term from reordering [7]. Furthermore, Eqs. (15) and (16) show additional
quantum effects due to the regularization ambiguity by the explicit appearance
of ${\bf F}$-dependent terms and highly divergent terms proportional to
$\delta^{2}(0)$ due to additional reordering.

\begin{center}
\bf{B. Angular momentum operator}
\end{center}

The classical angular momentum in our analysis is given by
\begin{eqnarray}
J&=&\int d^{2}x~( {\bf x} \times {\bf T}) \nonumber \\
 &=&\int d^{2}x~ \frac{1}{2i} {\bf x} \times
    \left[ \phi^{*} {\bf D} \phi-({\bf D} \phi)^{*} \phi \right] \nonumber \\
 &=& L + S,
\end{eqnarray}
where ${\bf T}^{i} \equiv T^{0i}$, the orbital angular momentum $L$ is
given by
\begin{eqnarray}
L =\int d^{2}x~ \frac{1}{2i} {\bf x} \times
     \left[ \phi^{*}\nabla \phi-(\nabla \phi^{*})\phi \right],
\end{eqnarray}
and $ S $ is the well-known Hagen's anomalous spin angular momentum [1,2,4]
\begin{eqnarray}
S= -\int d^{2} x~( {\bf x} \times {\bf A}) J^{0}= \frac{1}{4 \pi \kappa} Q^{2}.
\end{eqnarray}

Similar to the case of the Hamiltonian operator, four Hermitian different
operator
forms of angular momentum are also possible as follows
\begin{eqnarray}
J_{a} &\equiv&\int d^{2}x~ \frac{1}{2i} {\bf x} \times
    [ \phi^{\dagger} {\bf D}^{(1)} \phi-({\bf D}^{(1)} \phi)^{\dagger} \phi ]
    \nonumber \\
      &=& J+{\bf F} \times \int d^{2} x~ {\bf x} J^{0}, \nonumber \\
J_{b} &\equiv&\int d^{2}x~ \frac{1}{2i} {\bf x} \times
    [ \phi^{\dagger} {\bf D}^{(2)} \phi-({\bf D}^{(2)} \phi)^{\dagger} \phi ]
    \nonumber \\
      &=& J_{a} - {\bf F} \times \int d^{2} x~ {\bf x} J^{0}, \nonumber \\
J_{c} &\equiv&\int d^{2}x~ \frac{1}{2i} {\bf x} \times
    [  ({\bf D}^{(1)} \phi)\phi^{\dagger}-\phi ({\bf D}^{(1)} \phi)^{\dagger} ]
     \nonumber \\
      &=& J_{b} - \delta^{2} (0) \int d^{2} x~( {\bf x} \times {\bf A}),
       \nonumber \\
J_{d} &\equiv&\int d^{2}x~ \frac{1}{2i} {\bf x} \times
    [  ({\bf D}^{(2)} \phi)\phi^{\dagger}-\phi ({\bf D}^{(2)} \phi)^{\dagger} ]
    \nonumber \\
      &=& J_{c} - {\bf F} \times \int d^{2} x~ {\bf x} J^{0},
\end{eqnarray}
where $J =\int d^{2}x~(1/2i) {\bf x} \times
      \left[ \phi^{\dagger}\nabla \phi-(\nabla \phi^{\dagger})\phi \right]
      + (1/4 \pi \kappa)Q^{2}$.
Then the corresponding infinitesimal rotations of the matter field are
\begin{eqnarray}
\delta_{a} \phi &\equiv & [ J_{a}, \phi ({\bf x}) ] \nonumber \\
   &=&i {\bf x} \times \nabla \phi -\frac{Q}{2 \pi \kappa} \phi + ({\bf x}
       \times {\bf F}) \phi, \nonumber \\
\delta_{b} \phi &\equiv & [ J_{b}, \phi ({\bf x}) ] \nonumber \\
   &=&\delta_{a} \phi - ({\bf x} \times {\bf F}) \phi, \nonumber \\
\delta_{c} \phi &\equiv & [ J_{c}, \phi ({\bf x}) ] \nonumber \\
   &=&\delta_{b} \phi -\frac{\delta^{2} (0)}{\kappa}\int d^{2}x' {\bf x'}
    \cdot \overline {\bf \nabla}' {\cal D}({\bf x'-x}) \phi ({\bf x}) ,
    \nonumber \\
\delta_{d} \phi &\equiv & [ J_{d}, \phi ({\bf x}) ] \nonumber \\
   &=&\delta_{c} \phi - ({\bf x} \times {\bf F}) \phi.
\end{eqnarray}

Here we also see the iterative changes of the angular momentum operators and
their rotational anomalies due to the reordering and regularization ambiguity.

\begin{center}
\bf{C. Linear momentum and Galilei boost operators}
\end{center}

The classical linear momentum is given by
\begin{eqnarray}
{\bf P} &=&\int d^{2} x~ {\bf T} \nonumber \\
 &=&\int d^{2}x~ \frac{1}{2i}
     \left[ \phi^{*}\nabla \phi-(\nabla \phi^{*})\phi \right]
     -\int d^{2} x {\bf A} J^{0}  \nonumber \\
 &=&\int d^{2}x~ \frac{1}{2i}
     \left[ \phi^{*}\nabla \phi-(\nabla \phi^{*})\phi \right].
\end{eqnarray}
In the last step we used the fact $\int d^{2}x {\bf A}J^{0}$ vanishes due to
the symmetry property. However, this step would be illegitimate for the case of
point sources, i.e., sum of $\delta$ functions. This matter will be treated in
Sec. IV. The corresponding possible four Hermitian operator forms are
\begin{eqnarray}
{\bf P}_{a} &\equiv&\int d^{2}x~ \frac{1}{2i}
    [ \phi^{\dagger} {\bf D}^{(1)} \phi-({\bf D}^{(1)} \phi)^{\dagger} \phi ]
    \nonumber \\
      &=& {\bf P}- {\bf F}Q, \nonumber \\
{\bf P}_{b} &\equiv&\int d^{2}x~ \frac{1}{2i}
    [ \phi^{\dagger} {\bf D}^{(2)} \phi-({\bf D}^{(2)} \phi)^{\dagger} \phi ]
    \nonumber \\
      &=& {\bf P}_{a} + {\bf F} Q, \nonumber \\
{\bf P}_{c} &\equiv&\int d^{2}x~ \frac{1}{2i}
    [  ({\bf D}^{(1)} \phi)\phi^{\dagger}-\phi ({\bf D}^{(1)} \phi)^{\dagger} ]
    \nonumber \\
      &=& {\bf P}_{b} - \delta^{2} (0) \int d^{2} x {\bf A}, \nonumber \\
{\bf P}_{d} &\equiv&\int d^{2}x~ \frac{1}{2i}
    [  ({\bf D}^{(2)} \phi)\phi^{\dagger}-\phi ({\bf D}^{(2)} \phi)^{\dagger} ]
    \nonumber \\
      &=& {\bf P}_{c} + {\bf F}Q + \delta^{2} (0) {\bf F} V,
\end{eqnarray}
where ${\bf P}=\int dx^{2} (1/2i)[\phi^{\dagger}\nabla
\phi-(\nabla \phi^{\dagger})\phi]$, which is the usual momentum operator.
Although the anomalous terms in the Hamiltonian and angular momentum may not be
harmful in these cases because the anomaly can be attributed to some exotic
property of field itself, this is not the case for the linear momentum. In this
case, it can produce space-translationally noninvariant theory, which cannot
be allowed even for any exotic fields.

The corresponding four space-translational operations of the matter field are
\begin{eqnarray}
\left[ {\bf P}_{a}, \phi ({\bf x}) \right]
   &=&i \nabla \phi ({\bf x}) - {\bf F} \phi ({\bf x}),\nonumber \\
\left[ {\bf P}_{b}, \phi ({\bf x}) \right]
   &=&i \nabla \phi ({\bf x}),\nonumber  \\
\left[ {\bf P}_{c}, \phi ({\bf x}) \right]
   &=&i \nabla \phi ({\bf x})+\frac{\delta^{2} (0)}{\kappa}\int d^{2}x'
     \overline {\bf \nabla}' {\cal D}({\bf x-x'}) \phi ({\bf x}) ,\nonumber  \\
\left[ {\bf P}_{d}, \phi ({\bf x}) \right]
   &=&i \nabla \phi ({\bf x})+ {\bf F} \phi
       +\frac{\delta^{2} (0)}{\kappa}\int d^{2}x'
        \overline {\bf \nabla}' {\cal D}({\bf x-x'}) \phi ({\bf x})
\end{eqnarray}
showing the space-translational anomalies, which should vanish for the true
space-translational generators. We can classify true momentum operators
according to two possible values of
$\bf{F}$: i.e., ${\bf F}=0$ and ${\bf F} \neq 0$. For ${\bf F}=0$,
the true momentum operators are
\begin{eqnarray}
{\bf P}_{a}={\bf P}_{b} =
  \int d^{2}x~ \frac{1}{2i}
     [ \phi^{\dagger}\nabla \phi-(\nabla \phi^{\dagger})\phi ]={\bf P},
\end{eqnarray}
while, for ${\bf F}\neq 0$, they are
\begin{eqnarray}
{\bf P}_{b} =
\int d^{2}x~ \frac{1}{2i}
     [ \phi^{\dagger}\nabla \phi-(\nabla \phi^{\dagger})\phi ]={\bf P},
\end{eqnarray}
and all others cannot be accepted as the true momentum operators for the given
values of ${\bf F}$ due to wrong space-translational forms of the matter field.
This result means that the linear momentum operators can be considered as the
correct space-translational generators only with appropriate operator orderings
for a given regularization or quantization ambiguity ${\bf F}$. However
surprisingly, the final forms of linear momentum operators are all the same,
and
uniquely defined to be the canonical one which does not have the operator
ordering problem. Hence, the quantization ambiguity of the gauge-matter
commutation relation (11) does not induce the operator ordering ambiguity for
the linear momentum.

Now, by defining the Galilei boost operators corresponding to four momentum
operators of (23)
\begin{eqnarray}
{\bf G}_{k}&=&t {\bf P}_{k}- \int d^{2} x~ {\bf x} J_{0} ~~~~~(k=a,b,c,d),
\end{eqnarray}
the true boost operators become
\begin{eqnarray}
{\bf G}_{a}={\bf G}_{b}=t {\bf P}- \int d^{2} x~ {\bf x} J_{0}
\end{eqnarray}
for ${\bf F}=0$, and
\begin{eqnarray}
{\bf G}_{b}=t {\bf P}- \int d^{2} x~ {\bf x} J_{0}
\end{eqnarray}
for ${\bf F} \neq 0$. Furthermore, the final forms of these boost operators,
which are all the same, also have no operator ordering ambiguity.

\begin{center}
{\bf D. Galilei Covariance}
\end{center}

The Galilei covariance of the quantum field theory can be expressed by the
Galilei group
\begin{eqnarray}
&&[ P^{k}, P^{l} ]=0,~~ [ G^{k},G^{l} ]=0,~~ [ P^{k},G^{l} ]=i \delta_{kl} M,
 \nonumber \\
&&[ J,M ]=0, ~~[ H,H ]=0,~~[ G^{k}, M ]=0,~~[ P^{k},M ]=0, \nonumber \\
&&[ J,J ]=0, ~~[ P^{k},J ]=-i \epsilon_{km} P^{m}, ~~[ G^{k},J ]=
             -i \epsilon_{km} G^{m}, \nonumber \\
&&[ P^{k},H ]=0, ~~[ G^{k},H ]=-i P^{k}, ~~[ J,H ]=0,
\end{eqnarray}
where $ M$ is the mass operator. In this section we will show that this algebra
can be satisfied only for some particular orderings depending
the regularization
ambiguity ${\bf F}$. As a result, our model with inappropriate orderings
destroys
the Galilei covariance, which has been existed at the classical level
[7,11,12].
Although there is no {\it a priori} reason to reject the Galilei anomaly, we
only
consider the theory without this anomaly in order to determine the orderings
of the
generators uniquely.

First, by using the properly ordered momentum and Galilei boost operators of
Eqs. (24) and (28), we can easily confirm that the first three commutation
relations, i.e., ${\bf P}-{\bf P}$, ${\bf G}-{\bf G}$, ${\bf P}-{\bf G}$
relations, are satisfied with $ M=m \int d^{2} x J_{0}$.

Second, by considering the ${\bf P}-J $ relation, the commutation relations
for all possible orderings of angular momentum $ J _{a} \sim  J _{d}$ of
Eq.(21)
are obtained to be
\begin{eqnarray}
&&[P^{k}, J_{a}]=-i \epsilon _{km} P^{m} + {\bf F} \times \int d^{2} x~ {\bf x}
              \partial _{k} J_{0},\nonumber \\
&&[P^{k}, J_{b}]=-i \epsilon _{km} P^{m}, \nonumber \\
&&[P^{k}, J_{c}]=-i \epsilon _{km} P^{m} - \delta^{2} (0) \int d^{2} x~( {\bf
x}
            \times \partial _{k} {\bf A}),\nonumber \\
&&[P^{k}, J_{d}]=-i \epsilon _{km} P^{m} - {\bf F} \times \int d^{2} x~ {\bf x}
 \partial _{k} J_{0}  - \delta^{2} (0) \int d^{2} x~( {\bf x} \times
 \partial _{k} {\bf A}).
\end{eqnarray}
By noting that the anomalous term cannot vanish even if we assume highly
localized fields such that the appearing surface terms can be neglected, we see
that the appropriate orderings become
\begin{eqnarray}
J_{a}=J_{b}&=&
 \int d^{2}x~ \frac{1}{2i} {\bf x} \times
     [ \phi^{\dagger}\nabla \phi-(\nabla \phi^{\dagger})\phi ]
     +\frac{1}{4 \pi \kappa} Q^{2} =J
\end{eqnarray}
for ${\bf F}=0$, and
\begin{eqnarray}
J_{b}&=&
 \int d^{2}x~ \frac{1}{2i} {\bf x} \times
     [ \phi^{\dagger}\nabla \phi-(\nabla \phi^{\dagger})\phi ]
     +\frac{1}{4 \pi \kappa} Q^{2}=J
\end{eqnarray}
for ${\bf F }\neq 0$. This result shows that, although the appropriate
orderings of angular momentum are determined by the closure property of the
Galilei algebra, the final forms of the angular momentum operators are all
the same
as for the linear momentum operator. Although (33) and (34) contain the usual
anomalous spin term, we will use the representation of no rotational anomaly
such that we can treat the matter fields as the usual ones most safely in the
following. This is possible because the redefined angular momentum without the
anomalous spin term also satisfies the Galilei algebra [1,4].

Next, we consider the most nontrivial $J-H$ commutation relations. A lengthy
calculation shows that
\begin{eqnarray}
&&[J,H_{a}]=-\frac{i}{m} {\bf F} \times {\bf P},\nonumber \\
&&[J,H_{b}]=0,\nonumber \\
&&[J,H_{c}]=\frac{\delta^{2} (0)}{2m} \int d^{2}x \int d^{2} x' \int d^{2} x''
 [\overline \nabla {\cal D} ({\bf x-x'})]
 \cdot [\overline \nabla {\cal D} ({\bf x-x''})]
 [{\bf x''} \times \nabla '' J_{0}({\bf x''})] \nonumber \\
&&~~~~~~~~~~~\times J_{0}({\bf x''}), \nonumber \\
&&[J,H_{d}]=\frac{i}{m} {\bf F}\times {\bf P}
- \frac{\delta^{2} (0)}{m}{\bf F} \cdot \int d^{2}x \int d^{2} x'
 [\overline \nabla {\cal D} ({\bf x-x'})]
 [{\bf x'} \times \nabla ' J_{0}({\bf x'})] \nonumber \\
&&~~~+ \frac{\delta^{2} (0)}{2m} \int d^{2}x \int d^{2} x' \int d^{2} x''
 [\overline \nabla {\cal D} ({\bf x-x'})]
 \cdot [\overline \nabla {\cal D} ({\bf x-x''})]
 [{\bf x''} \times \nabla '' J_{0}({\bf x''})] J_{0}({\bf x''}).
\end{eqnarray}

Then the desired Hamiltonian becomes
\begin{eqnarray}
H_{a}=H_{b}= \int d^{2} x [\frac{1}{2m} (\nabla \phi^{ \dagger})
\cdot (\nabla \phi )-{\bf A} \cdot {\cal J} + \frac{1}{2m}{\bf A}^{2} J_{0} ]
\end{eqnarray}
for ${\bf F}=0$, and
\begin{eqnarray}
H_{b}=\int d^{2} x [\frac{1}{2m} (\nabla \phi^{ \dagger})
\cdot (\nabla \phi )-{\bf A} \cdot {\cal J}+\frac{1}{2m}{\bf A}^{2} J_{0} ]
\end{eqnarray}
for ${\bf F}\neq 0$. Here the matter current ${\cal J}$ is given by
\begin{eqnarray*}
{\cal J}=\frac{1}{2mi} [\phi^{ \dagger} \nabla \phi-(\nabla \phi^{ \dagger})
 \phi ].
\end{eqnarray*}
Although the final results of (36) and (37) are formally the same, the ordering
contents  are very different. (36) can be considered as the normal-ordered
Hamiltonian as
\begin{eqnarray}
H_{a}=~:H_{a}:~=~:H_{b}:~=H_{b}
\end{eqnarray}
due to the fact that ${\bf A}$ and $\phi($ or $\phi^{\dagger})$ commute in the
integration of (36) for the ${\bf F}=0$ case. However, this is not the case for
(37). Actually
\begin{eqnarray}
H_{b}=~:H_{b}: + \frac{1}{m} {\bf F} \cdot {\bf P}-\frac{1}{2m}{\bf F}^{2} Q
+\frac{1}{2m \kappa} \int d^{2} x \int d^{2} x' \phi^{ \dagger}({\bf x})
[{\overline \nabla} {\cal D}({\bf x-x'})]^{2}J_{0}({\bf x'})\phi({\bf x})
\end{eqnarray}
due to the fact that ${\bf A}$ and $\phi($ or $\phi^{\dagger})$, now, do not
commute in the integration of (36) for the ${\bf F}\neq 0$ case. Note that all
the last three terms of (39) cannot be simply subtracted for the Galilei
covariance because this cannot be recovered without some of them. Hence the
correct ordering of our theory is not the conventional normal ordering. If one
insists on conventional normal ordering, there is the Galilei anomaly for all
Hamiltonians $H_{a} \sim H_{d}$ for ${\bf F} \neq 0$. But, since this is
not compulsory, a rather more general ordering can be adopted, i.e.,
$ \phi^{\dagger} (\phi^{\dagger} \phi +\phi \phi^{\dagger} )\phi $,
$ i\phi^{\dagger} (\phi^{\dagger}\phi- \phi \phi^{\dagger} ) \phi , ...$,
which can be considered as a $modified~ normal~ ordering$. The modified normal
ordering satisfies all the necessary conditions of ordering such that properly
ordered generators should be Hermitian and annihilate the vacuum state.

Using the asserted form of $H$, ${\bf P}$, ${\bf G}$, and $J$, it is
straightforward to verify that all other commutation relations in Eq. (31) are
satisfied. Hence we showed that the operator ordering problem arisen from the
quantization ambiguity at the same point of $[{\bf A}({\bf x}), \phi({\bf x})]$
for the extended sources can be avoided by considering only the system with no
Galilei anomaly.
\\

\begin{center}
{\large \bf III. EQUATIONS OF MOTION FOR EXTENDED SOURCES }
\end{center}

We now study how the quantum equations of motion behave for our properly
ordered generators for the extended sources. To this end we define
\begin{eqnarray}
&&E^{i (1)} \equiv -\partial_{i}A^{0 (1)}-(\partial_{t}A^{i})^{(1)},
  \nonumber \\
&&E^{i (2)} \equiv -\partial_{i}A^{0 (2)}-(\partial_{t}A^{i})^{(2)},
\end{eqnarray}
where
\begin{eqnarray}
{A^{0}}^{(2)}({\bf x},t)&\equiv&-\frac{1}{\kappa} \int d^{2} x'{\bf J}^{(2)}
       ({\bf x'},t)
  \cdot \overline{{\nabla}}' \cal{D}({\bf x-x'}),\nonumber \\
{\bf J}^{(2)} &\equiv& \frac{1}{2mi} [\phi^{\dagger} {\bf D}^{(2)} \phi
         -({\bf D}^{(2)} \phi )^{\dagger} \phi ] \nonumber \\
              &=&{\bf J}^{(1)}+\frac{1}{m} {\bf F}J^{0}, \nonumber \\
 (\partial_{t}A^{i})^{(1)} &\equiv& i [H_{a}, A^{i} ],~~~
  (\partial_{t}A^{i})^{(2)} \equiv i [H_{b}, A^{i} ].
\end{eqnarray}
Then it is straightforward to show that
\begin{eqnarray}
B&=&-\frac{1}{\kappa} J^{0}, \nonumber \\
E^{i (1)}&=&\frac{1}{\kappa} \epsilon^{ij} J^{j (1)}, \nonumber \\
E^{i (2)}&=&\frac{1}{\kappa} \epsilon^{ij} J^{j (2)},
\end{eqnarray}
using
\begin{eqnarray}
&&(\partial_{t}J_{0})^{(1)} \equiv i[H_{a},J_{0}]=
       -\nabla \cdot {\bf J}^{(1)}, \nonumber \\
&&(\partial_{t}J_{0})^{(2)} \equiv i[H_{b},J_{0}]=
       -\nabla \cdot {\bf J}^{(2)}.
\end{eqnarray}

Now, Faraday's induction laws for each type become
\begin{eqnarray}
&&\nabla \times {\bf E}^{(1)}+(\partial_{t}B)^{(1)}=0, \nonumber \\
&&\nabla \times {\bf E}^{(2)}+(\partial_{t}B)^{(2)}=0.
\end{eqnarray}
Moreover, the equations of motion for the matter field for the second type in
Eq. (16) reduce to
\begin{eqnarray}
&&(i \partial_{t} \phi({\bf x}))^{(2)} \equiv
(i \partial_{t} \phi({\bf x}) )_{b} \nonumber \\
&&~~~~ =- \frac{1}{2m} { {\bf D}^{(2)} }^{2} \phi({\bf x})
    +{A^{0}}^{(2)}({\bf x}) \phi({\bf x})
   +  \frac{1}{2m {\kappa}^{2} }  \int d^{2} x'
    [\overline {\bf \nabla}' {\cal D}({\bf x'-x})]^{2} J_{0}({\bf x'})
     \phi({\bf x}).
\end{eqnarray}
All these results show that all the equations of the motion for both  gauge
and matter fields can be expressed as the similar form regardless of the types
of the orderings or the regularization prescriptions. This result strongly
suggests that the different orderings or regularization prescriptions do not
change the physics.

It seems appropriate to remark that there is also quantization ambiguity for
an $A^{0}-\phi$
commutation relation at the same point although this does not affect our
analysis:
\begin{eqnarray}
&&[A^{0(1)}({\bf x}),\phi({\bf x})] =
=- \frac{1}{i m }{\bf F} \cdot {\bf D}^{(1)} \phi ({\bf x}) -
\frac{1}{m \kappa^{2}} \int d^{2}x'' \overline \nabla {\cal D}({\bf x-x''})
\cdot \overline \nabla '' {\cal D}({\bf x-x''})J_{0}({\bf x''}) \phi ({\bf x}),
\nonumber \\
&&[A^{0(2)}({\bf x}),\phi({\bf x})] =
[A^{0(1)}({\bf x}),\phi({\bf x})] -
\frac{1}{m \kappa}{\bf F}^{2} \phi({\bf x}).
\end{eqnarray}
\\

\begin{center}
{\large \bf IV. ORDERING PROBLEM FOR POINT SOURCES}
\end{center}

For the point source system much more care is needed for nonzero ${\bf F}$.
This is essentially due to the fact that integrals such as $\int d^{2}x
{\bf A}J^{0}$
do not vanish for the point sources, i.e., the sum of $\delta$ functions.
In this
case it becomes
\begin{eqnarray}
\int d^{2}x ~{\bf A}J_{0}=-({\bf F}/ { 2 \pi}) \int d^{2} x J_{0}^{2},
\end{eqnarray}
which does not vanish for a nonzero ${\bf F}$. Then the linear momentum
operators
of (24) are modified as
\begin{eqnarray}
{\tilde{\bf P}}_{k}={\bf P}_{k}-({\bf F}/ { 2 \pi}) \int d^{2} x J_{0}^{2}
 ~~~(k=a,b,c,d).
\end{eqnarray}
However, one can easily find that none of these operators can be considered as
the true momentum operators for nonzero ${\bf F}$ by considering the
space-translational operations of the matter field:
\begin{eqnarray}
\left[ {\tilde{\bf P}}_{a}, \phi ({\bf x}) \right]
   &=&i \nabla \phi ({\bf x}) - {\bf F} \phi ({\bf x})+\frac{{\bf F}}{2 \pi}
        J^{0}({\bf x}) \phi({\bf x}),\nonumber \\
\left[ {\tilde{\bf P}}_{b}, \phi ({\bf x}) \right]
   &=&i \nabla \phi ({\bf x})+\frac{{\bf F}}{2 \pi}J^{0}({\bf x}) \phi({\bf
x}),
           \nonumber  \\
\left[ {\tilde{\bf P}}_{c}, \phi ({\bf x}) \right]
   &=&i \nabla \phi ({\bf x})+\frac{\delta^{2} (0)}{\kappa}\int d^{2}x'
     \overline {\bf \nabla}' {\cal D}({\bf x-x'}) \phi ({\bf x})+\frac{{\bf F}}
           {2 \pi}J^{0}({\bf x}) \phi({\bf x}) ,\nonumber  \\
\left[ {\tilde{\bf P}}_{d}, \phi ({\bf x}) \right]
   &=&i \nabla \phi ({\bf x})+ {\bf F} \phi
       +\frac{\delta^{2} (0)}{\kappa}\int d^{2}x'
        \overline {\bf \nabla}' {\cal D}({\bf x-x'}) \phi ({\bf x})
           +\frac{{\bf F}}{2 \pi}J^{0}({\bf x}) \phi({\bf x}).
\end{eqnarray}
Any operators in ${\tilde{\bf P}}_{a} \sim {\tilde{\bf P}}_{d}$ do not produce
the correct space translation for a nonzero ${\bf F}$. Since the regularization
prescription of a nonzero ${\bf F}$ is not physically allowed for the above
reason, the usual prescription of ${\bf F}=0$ of Refs. [1], [4], and [7]
is the only
possible one. This is in sharp contrast to the case of extended source system
where any regularization prescriptions are allowed although the proper
orderings
are determined depending on the prescriptions. Although a similar analysis
may be
performed for other generators, these are redundant ones because the analysis
for the linear momentum operator gives the most restrictive condition for the
regularization ambiguity ${\bf F}$ already.

Now we discuss the physical implications of this result. It is well known that
the nonrelativistic Abelian Chern-Simons gauge theory does not exhibit
self-energy at the classical level [7,14]. This is crucially due to the exact
cancellation of electric and magnetic field contributions to the Lorentz force
by the classical equations motion (3) and (4). By noting that
$[{\bf A}({\bf x}), \phi({\bf x})]  = 0 $ implies that there
is no self-interactions even at the quantum level in our model, one can expect
that the usual prescription of ${\bf F}=0$ is the only consistent one with the
classical results. Actually this can be easily confirmed by the fact that the
Schr\"{o}dinger equation for nonzero {\bf F} reveals the self-interaction
induced by quantum corrections. Explicit manipulations for the one-body
Schr\"{o}dinger equation give the relations
\begin{eqnarray}
&&E u^{a}_{E}({\bf x})=-\frac{1}{2m} \nabla^{2} u^{a}_{E}({\bf x}),\nonumber \\
&&E u^{b}_{E}({\bf x})=-\frac{1}{2m} (\nabla+i{\bf F})^{2} u^{b}_{E}({\bf x}),
     \nonumber \\
&&E u^{c}_{E}({\bf x})=[-\frac{1}{2m} (\nabla+i{\bf F})^{2}
+ \frac{\delta^{2} (0)}{2m \kappa^{2} } \int d^{2} x'[\overline {\bf \nabla}'
  {\cal D}({\bf x'-x})]^{2} ] u^{c}_{E}({\bf x}), \nonumber \\
&&E u^{d}_{E}({\bf x})=[-\frac{1}{2m} (\nabla+2i{\bf F})^{2}-\frac{1}{m}
{\bf F}^{2}+ \frac{\delta^{2} (0)}{2m \kappa^{2} } \int d^{2} x'
    [\overline {\bf \nabla}' {\cal D}({\bf x'-x})]^{2} \nonumber \\
&&~~~~~~~~~~~~~~~~~~~ - \frac{\delta^{2} (0)}{m \kappa } {\bf F} \cdot
 \int d^{2}x'\overline {\bf \nabla}' {\cal D}({\bf x'-x}) ] u^{d}_{E}({\bf x}).
\end{eqnarray}
In deriving these equations we use the relations
\begin{eqnarray}
u_{E}({\bf x}_{1},...,{\bf x}_{N})=<0 \mid \phi({\bf x})...\phi({\bf x}_{N})
 \mid E,N > \nonumber \\
E u_{E}({\bf x}_{1},...,{\bf x}_{N})=<0 \mid [\phi({\bf x})...\phi({\bf
x}_{N}),
  H] \mid E,N >
\end{eqnarray}
with the energy and particle number eigenstate $\mid E,N>$ and the vacuum state
$\mid 0 >$ satisfying
\begin{eqnarray}
&&H \mid E,N >=E \mid E,N >,\nonumber \\
&&N \mid E,N >=N \mid E,N >,\nonumber \\
&&<0 \mid \phi^{\dagger}({\bf x})=\phi({\bf x}) \mid 0 >=0, \nonumber \\
&&H \mid 0 >=N \mid 0>=0
\end{eqnarray}
for each Hamiltonian of (15) and we use the equations of motion (16), which are
valid for the case of the point sources also. The result (50) explicitly shows
that only the usual prescription of ${\bf F}=0$ exhibits no self-interaction
with $u^{a}_{E}$ and $u^{b}_{E}$, which are the same in this prescription.
Note that, although $N$-body Schr\"odinger equation [1,7,14] may be analyzed
generally, this does not change the essence of our argument. As a result, we
recognize that the usual ordering corresponding to type-$a$ and type-$b$,
which are
the same for ${\bf F}=0$, is the only possible one due to the unique
determination of ${\bf F}=0$ for the consistency, i.e., the space-translational
invariance of the model at the quantum level.
\\

\begin{center}
{\large \bf V. CONCLUSION}
\end{center}

In this paper we have shown that the nontrivial operator ordering problem of
the nonrelativistic Abelian Chern-Simons theory in the Coulomb gauge can be
avoided if we require Galilei covariance even at the quantum level or the
consistency. The requirement of Galilei covariance is nontrivial because we
do not have any principle to disregard the Galilei anomaly of the model in
$D=2+1$ dimension.

Actually, the recovery of the covariance for the extended sources is only
possible when we choose some specific orderings, which cannot be the same as
the well-known normal ordering but still satisfy all the necessary conditions
at the same point of the proper orderings for the unusual prescription
$[{\bf A}({\bf x}), \phi({\bf x})]  \neq 0 $. These specific
orderings are the same as the usual ordering for the usual prescription
$[{\bf A}({\bf x}), \phi({\bf x})]  = 0 $. However, we have
shown that both the usual and unusual orderings or regularization prescriptions
describe the same physics by noting that all the equations of motion can
express the same form regardless of the types of the orderings or
regularization prescriptions.

On the other hand, for the point source system, the requirement of consistency,
i.e., space-translational invariance, which can be guaranteed by the proper
momentum operator, can be satisfied only for the usual prescription and not for
the unusual ones. Hence only the orderings, which are equivalent to the usual
one, are allowed in this case. Moreover, only for this usual ordering or
prescription is the quantum theory consistent with the classical results.

As final remarks, we first note that our analysis for the extended sources
may be useful for the relativistic Chern-Simons theory since the sources of the
relativistic system are inherently extended although the anyonicity of
the model
is still debatable. Second, we note that our results are not changed even when
the usual quadratic self-interaction term for the matter field are introduced
although it may change the structure of the conformal group [7,11,12,15].
Finally, although we only considered the representation without rotational
anomaly, it is questionable whether a similar result can be also obtained
for the
representation with rotational anomaly.
\\

\begin{center}
{\large \bf ACKNOWLEDGMENTS}
\end{center}

We thank H. Min and W.T. Kim for helpful discussions.
The present work was supported in part by the Basic Science Research Institute
program, Ministry of Education, Project No. 94-2414, and the Korea Science
and Engineering Foundation through the Center for Theoretical Physics.

\newpage
$^{a}$ Electronic address: mipark@physics.sogang.ac.kr

$^{b}$ Electronic address: yjpark@ccs.sogang.ac.kr

\begin{description}
\item{[1]} C. R. Hagen, Phys. Rev. {\bf D 31}, 848 (1985).
\item{[2]} C. R. Hagen, Phys. Rev. {\bf D 31}, 2135 (1985);
D. Arovas, J. R. Schrieffer, F. Wilczek, and A. Zee, Nucl. Phys. {\bf B251},
117 (1985); J. Fr\"{o}hlich and P.-A. Marchetti, Lett. Math. Phys. {\bf 16},
347 (1988); A. M. Polyakov, Mod. Phys. Lett. {\bf A3}, 325 (1988).
\item{[3]} G. W. Semenoff, Phys. Rev. Lett. {\bf 61}, 517 (1988).
\item{[4]} C. R. Hagen, Ann. Phys. (N.Y.) {\bf 157}, 342 (1984);
Phys. Rev. {\bf D 31}, 331 (1985).
\item{[5]} J.F. Schonfeld, Nucl. Phys. {\bf B185}, 157 (1981);
S. Deser, R. Jackiw, and S. Templeton, Phys. Rev. Lett. {\bf 48}, 975 (1982);
Ann. Phys. (N.Y.) {\bf 140}, 372 (1982); S. Deser and R. Jackiw, Phys. Lett.
{\bf B139}, 371 (1984); D. Boyanovsky, R. Blankenbecler, and R.
Yabalin, Nucl. Phys. {\bf B270}, 483 (1986).
\item{[6]} G. V. Dunne, R. Jackiw, and C. A. Trugenberger, Ann. Phys. (N.Y.)
{\bf 194}, 197 (1989); C.A. Trugenberger, R. Menikoff, and D.H. Sharp,
Phys. Rev. Lett. {\bf 67}, 1922 (1991); C.A. Trugenberger, Phys. Rev.
{\bf D 45},
3807 (1992).
\item{[7]} R. Jackiw and S.-Y. Pi, Phys. Rev. {\bf D 42}, 3500 (1900),
 {\bf 48}, 3929 ({\bf E}) (1993).
\item{[8]} P. A. M. Dirac, {\it Lecture on Quantum Mechanics} (Belfer Graduate
School of Science, Yeshiva University Press, New York, 1964); P. Gerbert,
Phys. Rev.
{\bf D 42}, 543 (1990); R. Banerjee, Phys. Rev. Lett. {\bf 69}, 17 (1992).
\item{[9]} L. Faddeev and R. Jackiw, Phys. Rev. Lett. {\bf 60}, 1692 (1988);
  T. Matsuyama, Phys. Lett. {\bf B228}, 99 (1989).
\item{[10]} This can be explicitly confirmed by considering the commutation of
Eq.(11) with presumed correct momentum and Hamiltonian operators.
\item{[11]} M. Leblanc, G. Lozano and H. Min, Ann. Phys. (N.Y.) {\bf 219},
328 (1992).
\item{[12]} R. Jackiw, Ann. Phys. (N.Y.) {\bf 201}, 83 (1990).
\item{[13]} P. Gerbert [8].
\item{[14]} C. Kim, C. Lee, P. Ko, B.-H. Lee, and H. Min, Phys. Rev. {\bf D
48},
1821 (1993).
\item{[15]} C.R. Hagen, Phys. Rev. {\bf D 5}, 377 (1972); R. Jackiw
and S.-Y. Pi,
Phys. Rev. Lett. {\bf 64}, 2969 (1990); D.Z. Freedman, G. Lozano, and N. Rius,
Phys. Rev. {\bf D 49}, 1054 (1994); O. Bergman and G. Lozano, Ann. Phys. (N.Y.)
{\bf 229}, 416 (1994); D. Bak and O. Bergman, MIT preprint(1994) CTP No. 2283
(hep-th/9403104).

\end{description}

\end{document}